\documentclass[11pt,twoside]{article}
\usepackage{asp2010}

\begin{document}

\title{The Interstellar Reddening Law within 3kpc from the Sun}
\author{Hwankyung Sung,$^1$ and M. S. Bessell$^2$}
\affil{$^1$Department of Astronomy and Space Science, Sejong University, 209 Neungdongro, Kwangjin-gu, Seoul 143-747, Korea}
\affil{$^2$Research School of Astronomy and Astrophysics, Australian
National University, MSO, Cotter Road, Weston, ACT 2611, Australia}

\begin{abstract}
We have investigated the interstellar reddening law of young open
clusters within 3kpc from the Sun using optical, near-IR 2MASS, and 
{\it Spitzer} IRAC data. The total-to-selective extinction
ratio $R_V$ of 162 young open clusters ($\log ~t_{age} \lesssim 
7.3$) listed in the open cluster database WEBDA is determined
from the color excess ratios. The young open clusters
in the Sgr-Car arm show a relatively higher $R_V$, those in
the Per arm and in the Cygnus region of the local arm show a relatively
smaller value, and those in the Mon-CMa region of the local arm show
a normal value ($R_V \approx 3.1$).

\end{abstract}

\section{Introduction}

To determine the physical parameters of stars and clusters accurately, we should
know the correct interstellar reddening. Precise knowledge of $R_V$ is of crucial 
importance for the determination of distances of reddened stars or clusters.
Many investigators determine
the total extinction $A_V$ by assuming the total-to-selective extinction
ratio $R_V$ to be 3.1. In 1940s, from multicolor photometry of many O and B type stars
\citet{sw43} concluded that the law of selective absorption is the same for all directions
in the Galaxy. Later \citet{jb63} conducted multi-color photometry of O, B stars and
showed that there is no unique reddening law. In addition, they showed that there is
a minor variation of $R_V$ with galactic longitude. Later \citet{serkowski75} confirmed
such a variation from the spatial variation of wavelength of maximum polarization.
\citet{whittet77} presented a functional form of $R_V$ variation against galactic
longitude. Recently \citet{fitzpatrick09} showed that the interstellar reddening law $R_V$ is
different for different sightlines.

Recently \citet{hur12} showed that an abnormal reddenig law for
the intracluster medium of the young open clusters in the $\eta$ Carina nebula.
However they obtained a fairly normal reddening law for the foreground stars.
On the other hand, \citet{sung13a} found a nearly normal reddening law toward
the young open cluster NGC 6231 from the optical to mid-IR photometric data. 
Now it is possible to study the reddening law toward young open clusters
in a homogeneous way because the homogeneous near-IR 2MASS data as well as
mid-IR {\it Spitzer} IRAC images are available for many young open clusters.
In this paper we present a preliminary result from our systematic investigation of the reddening law.

\begin{figure*}[!t]
\begin{center}
\plotone{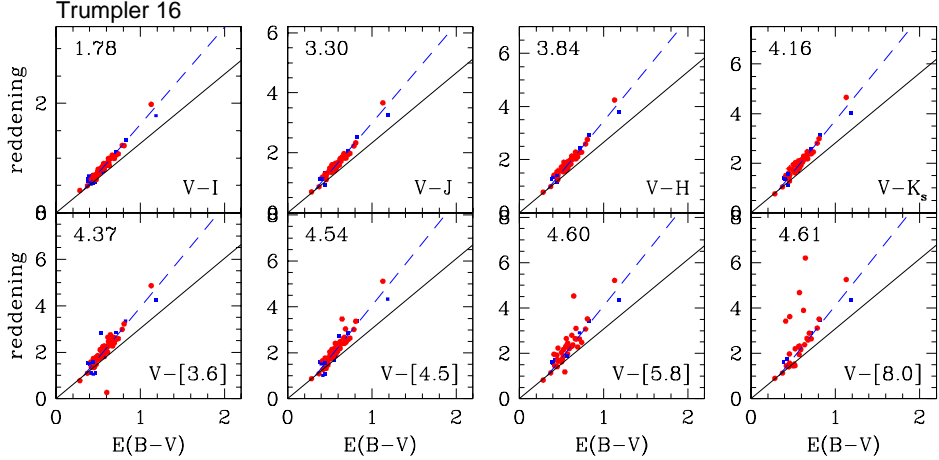}
\caption{Reddening law of Trumpler 16 from the optical to mid-IR. The reddening for a given color
with respect to $E(B-V)$ is shown. The optical and near-IR data are from \citet{hur12} and
2MASS, respectively, while mid-IR photometry was obtained from the {\it Spitzer} IRAC images.
The solid and dashed lines represent the normal reddening law for the foreground ($R_{V,fg} = 3.1$)
and the abnormal reddening law for the intracluster medium of Trumpler 16 region ($R_{V,cl} = 4.6$),
respectively. Large dots and squares denote stars with higher proper motion membership probability
($P_\mu \geq 70$) and with lower probability ($P_\mu < 30$), respectively.
\label{fig1} }
\end{center}
\end{figure*}

\section{Large Scale Variation of $R_V$}

The reddening law for the young open clusters is determined as the same way as the method described
in \citet{hur12,sung13b}. The intrinsic color relations between ($B-V$)$_0$ and ($V-\lambda$)$_0$
for 2MASS $JHK_s$ bands are presented in \citet{sung13b}, and those for {\it Spitzer} IRAC bands
will be published in the forthcoming paper.

We have analyzed the photometric data for 192 young open clusters with $\log \tau_{age} \lesssim
7.3$. Among them 16 clusters (e.g. Mayer 1, Be 65, Cz 13, Cr 173, etc) are either lack of $UBV$ 
photoelectric or CCD data or ($U-B$) color. And 4 clusters (e.g. Cz 20, NGC 2453, etc) are not young 
open clusters, 6 clusters (Cr 96, Cr 107, NGC 2448, etc) seem to be not clusters, and 4 clusters (BH 121, BH 205, 
IC 2948, and Loden 153) are parts of the other young clusters. And therefore we determined the reddening law of
162 young open clusters. We also determined the color excess ratios in the {\it Spitzer} IRAC 
bands for 38 clusters. The $R_V$ of the young open clusters is determined the relation between
$R_V$ and color excess ratios by \citet{gv89}.
Figure \ref{fig1} shows the color excess ratios for the OB stars in the young open cluster
Trumpler 16. A few foreground B type stars show a fairly normal reddening law of $R_{V,fg} = 3.1$,
while cluster OB stars behave abnormally and well match to the solid line of $R_{V,cl} = 4.6$.
Although most stars do not show any excess emission in the optical to near-IR, some
stars show an evident excess emission in the mid-IR, especially in 8.0 $\mu m$.

\begin{figure*}[!t]
\begin{center}
\plotone{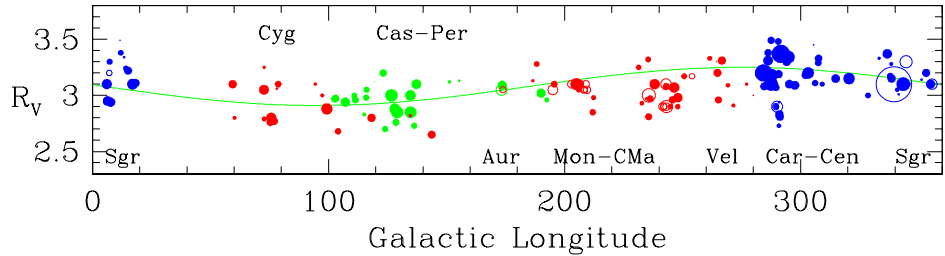}
\caption{Plot of $R_V$ against galactic longitude $l$. The solid curve is the variation of
$R_V$ from \citet{whittet77}. Large dots represent the $R_V$ determined from the color excess ratios,
while open circles denote the $R_V$ determined from less reliable data.
The size of symbols is proportional to the number of OB stars used in the determination
of $R_V$.  \label{fig2} }
\end{center}
\end{figure*}

In this paper we only describe the reddening law of the general interstellar medium.
And therefore we only take the $R_{V,fg}$ for the young open clusters with
an abnormal $R_{V,cl}$. Figure \ref{fig2} shows the $R_{V}$
with respect to galactic longitude. For some clusters we cannot determine the $R_V$
reliably because of small values of reddening or no early type stars in the foreground.
The young open clusters in the Sgr-Car arm show a relatively high value ($R_V \approx 3.2$), those in
the Per arm and in the Cygnus region have a smaller value ($R_V \approx 2.9$), and those
in the Mon-CMa regions have a nearly normal value ($R_V \approx 3.1$).
In addition the fluctuation of $R_V$ is also noticeable ($\sigma_{R_V} \approx 0.15$).
However the overall variation of $R_V$ well follows the equation given by \citet{whittet77}
except for the CMa-Vel direction.

\section{Conclusion}

From the analysis of photometric data for 162 young open clusters
we confirmed the large scale variation of the reddening law in the galactic plane,
which implies the real variation of dust size distribution in the galactic plane.
We also found a large fluctuation of $R_V$ for a given region. The cause of this fluctuation
is still uncertain, and therefore a systematic photometric survey of open clusters
to provide homogeneous photometric data, e.g.
the Sejong Open cluster Survey (SOS) \citep{sung13b} is required to confirm the real
fluctuation.

\acknowledgements H.Sung acknowledges the support of the National Research Foundation of Korea (NRF)
funded by the Korea Government (MEST) (Grant No.2013031015).

{}

\end{document}